\documentclass[12pt]{iopart}
\usepackage{epsfig}
\begin{document}

\title[]{Stable quark stars beyond neutron stars:
can they account for the missing matter ?}

\author{Peter Minkowski and Sonja Kabana\ddag
}

\address{\dag\ 
Institute for Theoretical Physics, University of Bern, Sidlerstrasse 5, 3012 Bern, Switzerland\\
	E-mail: mink@itp.unibe.ch}

\address{\ddag\ 
Laboratory for High Energy Physics, University of Bern, 
        Sidlerstrasse 5, 3012 Bern, Switzerland\\
	E-mail: sonja.kabana@cern.ch}




% \preprint{BUHE-01-?? and ... \hepth{9912999}}	% OR: \preprint{Aaaa/Mm/Yy\\Aaa-aa/Nnnnnn}
			  	% Use \hepth etc. also in bibliography.  

\begin{abstract}
The structure of a spherically symmetric stable
dark 'star' is discussed, at zero temperature, containing
1) a core of quarks in the deconfined phase
and antileptons
2) a shell of hadrons in particular $n$, $p$, $\Lambda$ and $\Sigma^-$
and leptons or antileptons
and 3) a shell of hydrogen in the superfluid phase.
\noindent
If the superfluid hydrogen phase goes over into the electromagnetic plasma phase
at densities well below one atom / $( \ 10 \ fm \ )^{\ 3}$, as is usually assumed,
the hydrogen shell is insignificant for the mass and the radius
of the 'star'. These quantities are then determined approximatively :
mass = 1.8 solar masses and
radius = 9.2 km.
\noindent
On the contrary if densities of the order of one atom / $( \ 10 \ fm \ )^{\ 3}$
do form a stable hydrogen superfluid phase,
we find a large range of possible masses from 1.8  to
375 solar masses.  The radii vary accordingly from 9 to 1200 km.
\end{abstract}

% \conference{Strange Quark Matter 2001}

%\keywords{Dark Matter, Quark stars}

%\maketitle  IS IGNORED %%%%%%%%%%%

\vspace*{-0.5cm}
\vspace*{-0.5cm}

\section{Introduction}

\noindent
Many authors have found solutions for strange quark stars
with shells of hadronic composition
\cite{astrostrangelets1}.
We investigate here the structure of a spherically symmetric stable
dark 'star' as outlined in the abstract.
We use for the quark core the Tolman solution VI \cite{tolman} with infinite
central energy density and pressure.
We also introduce a superfluid hydrogen shell which can be very large
depending on its inner energy density.
For normal hydrogen densities the hydrogen shell is very thin
and negligible in mass.
Objects with a mass of few solar masses can in principle be
detected by gravitational lensing.
However, in case of abnormally large initial energy density
of the hydrogen shell, we arrive at very large masses of the star,
which can not be easily detected by lensing.
In this case, such stars are candidates for 'nonbaryonic' dark matter.
\noindent
The three shells are discussed in the following section.
For details of this analysis see reference \cite{issi}.

\begin{figure}[htb]
\begin{center}
%\hskip 20mm
%\epsfig{file=early_T_vs_mub_rot_lin3.eps,width=7.0cm} 
\epsfig{file=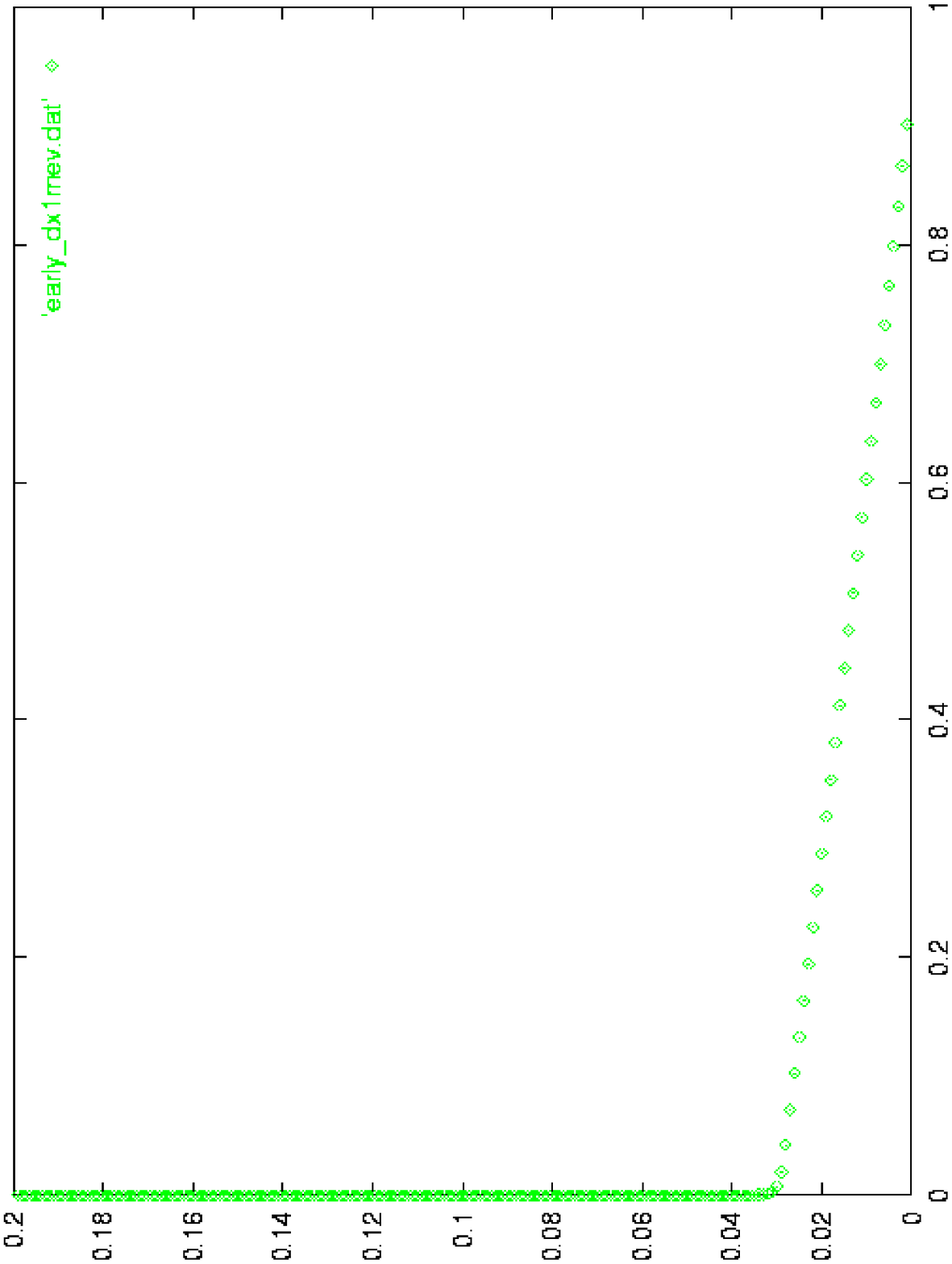,width=9.0cm} 
\caption{\label{t} 
\hspace*{-0.5cm} \\
\hspace*{-0.5cm} 
\begin{tabular}{@{\hspace*{-0.0cm}}l}
\\  Temperature in GeV  as a function of the baryochemical  potential
$\mu_B$ in GeV, \\ showing the path followed by the early universe  
after the QCD phase 
transition.
\end{tabular}}
\end{center}
\end{figure}
\begin{figure}[h]
\begin{center}
\epsfig{file=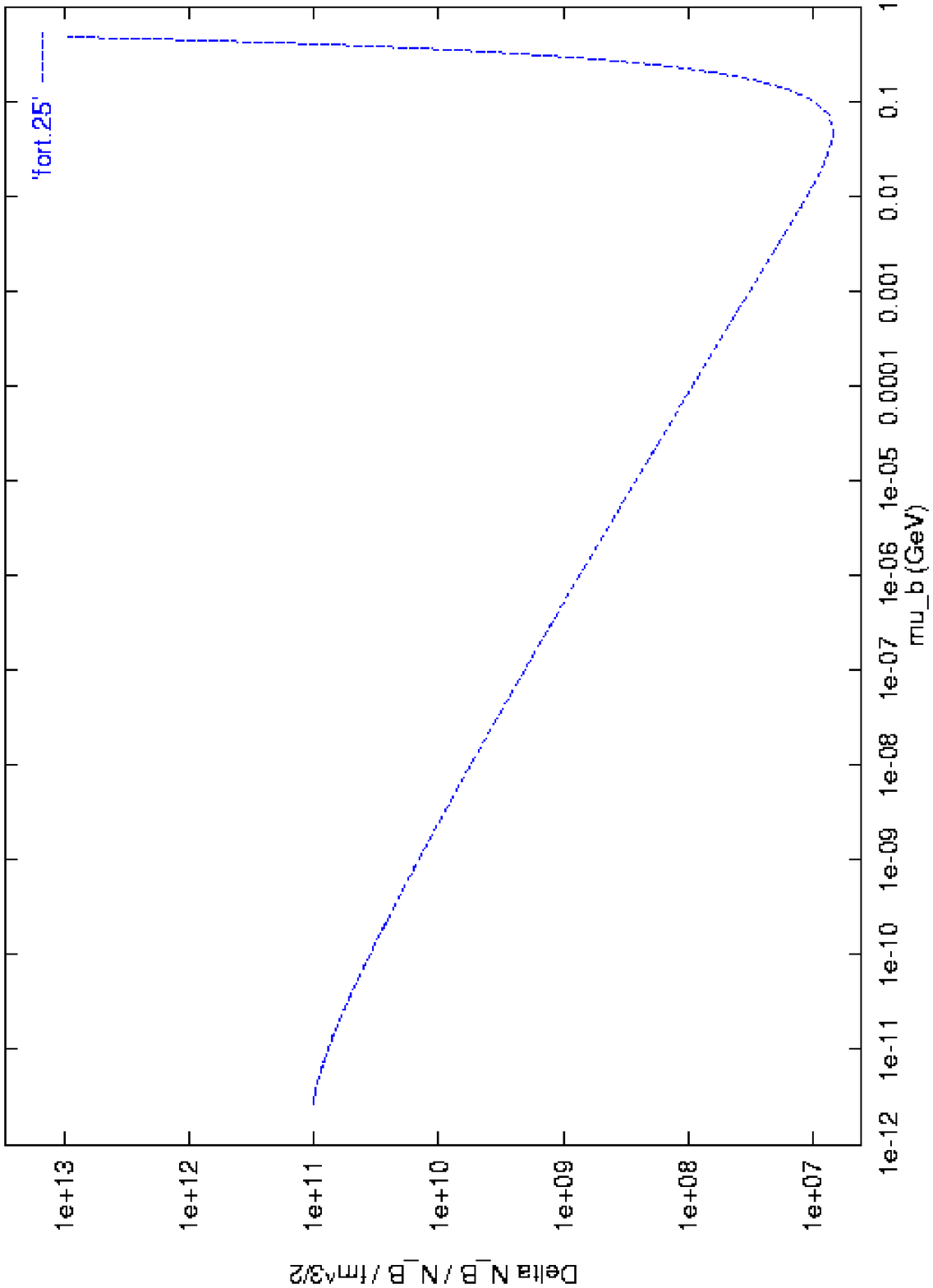,width=9.0cm} 
\caption{\label{f}  \\
\begin{tabular}{@{\hspace*{0.0cm}}l}
Relative fluctuations of the net baryon number normalized to  
a volume of 1 $fm^3$ \\ as a function of  the baryochemical potential, 
following the path 
of figure 1
\end{tabular}}
\end{center}
\end{figure}

\vspace*{-0.2cm}

\section{The shells of the star}

\vspace*{-0.3cm}

We consider 3 shells of radii r0 (quark shell), r1 (hadron shell), and r2
(hydrogen shell).
The quark core of the star is described according to Tolman's
solution VI \cite{tolman} and 
energy density and pressure are given by
\\
$ \ \rho_e \ = \frac{3} {56 \ \pi \ G_N} \frac{1} {r^2} \ $
and
$ \ p \ = \frac{1} {56 \ \pi \ G_N} \frac{1} {r^2}
\frac {(1 - 9 \ (B/A) \ r) }
{(1 - (B/A) \ r)} \ $.
\\
Where A, B are integration constants.
We consider the core made up of u,d,s quarks and antileptons.
allowing for the three chemical potentials associated with charge ($\mu_Q$),
baryon number ($\mu_B$) and  total lepton number ($\mu_L$).
The hadronic phase is composed of baryons, leptons and antileptons.
%********************* corrections
We use in the hadronic phase the vacuum pressure from \cite{mapping}.
%********************* new 26.02.2002
The TOV equations allow for a cosmological term, which in the hadronic
phases must be chosen to cancel the positive vacuum pressure.
It thus appears with negative sign in the deconfined
core phase. The Gibbs equal pressure condition cannot be met,
at zero temperature, neglecting
a repulsive and very large second virial coefficient in the hadronic phase.
Thus the equilibrium condition corresponds to equal energy density \cite{issi}.
%*********************8 new 26.02.2002
The hydrodynamic and QCD equilibrium conditions between the two phases
(hadronic and plasma) together with the condition of
%********************* new 26.02.2002
locally 
%********************* new 26.02.2002
vanishing overall electric charge determine
the core radius, mass
 and all chemical potentials for charge, baryon number and
lepton number.
We derive the mass $M$ and the radius $r0$ of the core:
M(quark core) = 0.457 solar masses
and r0(quark core) =  3.149   km. 
The baryon number of the core is
$N_{\ B \ core} \ \sim \ 0.287 \ N_{\ B \ sun}$.
%********************  corrections 
Therefore the core itself is not bound, 
but  stabilized by the outer hadron shell.
\noindent
We simplify the hadron shell composition structure by considering
only neutrons, protons and electrons, though near the QCD transition zone
 also
$\Sigma^-$, $\Lambda$, $\mu^+$, $e^+$ and all antineutrinos intervene.
We obtain : 
$ M \ ( \ \mbox{hadronic shell} \ ) \ =  \ 1.344 \ \mbox{solar masses}$
and
$r1 \ ( \ \mbox{hadronic shell} \ ) = 9.047 \ $ km.
Therefore the total mass of both phases
 inside the radius r1 is:
$M \ ( \ \mbox{total inside r1} \ ) \ =  \ 1.801 \ \mbox{solar masses}$
and the baryon number inside r=r1 is 
1.853 $N_B(sun)$.
The nonrelativistic approximation of the TOV
equation used here overestimates somewhat both mass and radius.
\noindent
We proceed to calculate the mass and radius of the hydrogen shell
which starts at radius r1 and ends at radius r2.
We use here the original approximate relations for superfluid
hydrogen
\footnote{We do not include any further light elements, fermions or bosons,
in the hydrogen shell.},
or alike boson-condensed non relativistic system,
first described by N. N. Bogolubov \cite{bogolubov}.
The results dependent on the parameter H, which is the
 initial energy density of hydrogen at r=r1, normalized to
$ \alpha m_e m_p m_H^2$.
The total radii, masses and baryon numbers are shown in table 1
together with H.
The normal hydrogen shell corresponds to the first line.

\vspace*{-0.5cm}
\begin{center}
\begin{table}[htb]
\begin{tabular}{|c|c|c|c|}
 \hline
$H$      &    Radius of dark 'star' & Mass of dark 'star'  & $N_B$ \\
         &    (km)                 &  (solar masses)     &   ($N_B(sun)$) \\
 \hline
5 $10^{-3}$   &       9.204           &     1.801   & 1.853
\\ \hline
0.290   &       411       &     2.191          & 2.243
\\ \hline
0.295    &    674       &       3.179          & 3.232
\\ \hline
0.310    &   1061       &       12.63           & 12.68
\\ \hline
0.50     &  1203        &      168.13           & 168.18
\\ \hline
0.625     &  1208               & 271.06        & 271.11
\\ \hline
0.75     &  1210        &       374.02          & 374.07
\\ \hline
\end{tabular}
\caption{Mass and radius for the complete 'star' with a quark (anti)lepton core,
a hadron shell and a hydrogen shell. H is the energy density of the hydrogen shell
at r=r1 normalised to $\rho_{e1} = \alpha m_e m_p m_H^2$.}
\label{mass}
\vspace*{-0.5cm}
\end{table}
\end{center}

%\vspace*{-0.5cm}
\vspace*{-0.5cm}

\section{The big annihilation}
\vspace*{-0.3cm}

We follow the path of the early universe right after the QCD phase transition
which is important for the genesis of quark stars.
In particular fig. \ref{t} shows the temperature as a function
of the baryochemical potential. 
At about T $\sim$ 30 MeV the annihilation of 
matter and antimatter begins.
We denote this 'the big annihilation'.
Note that laboratory experiments measured a copious production
of antibaryons at a temperature of $\sim$ 170 MeV
see e.g. \cite{na52}.
The statistical fluctuations of net baryon number normalized
to a volume of 1 fm$^3$  following this path, are
shown in fig. \ref{f}. They show maxima at the phase boundary
and at temperatures below 10 MeV.
 The rest of the condensation of such a star
needs large gravitational fluctuations too.

\vspace*{-0.3cm}

\section{Conclusions}
\vspace*{-0.3cm}

\noindent
We investigated the structure of a spherically symmetric stable
dark 'star'.
For a low density hydrogen shell 
we find for the mass and the radius
of the 'star'
$ \mbox{mass} \ = \ 1.8 \ \mbox{solar masses}$
and $ \mbox{radius} \ = \ 9.2 \ \mbox{km}$.
\noindent
For a high density hydrogen shell supporting densities of
up to one atom / $( \ 10 \ fm \ )^{\ 3}$
we find a large range of possible masses from 1.8  to
375 solar masses.  The radii vary accordingly from 9 to 1200 km.
\noindent
Such objects are candidates for dark matter
of any kind, despite their clearly baryonic nature,
if  they have been formed in such a way that
they do not affect nucleosynthesis
and the cosmic microwave background fluctuations
at decoupling.
We have calculated the path of the early universe after the QCD phase transition
in the (T, $\mu_B$) plane,
and the net baryon number fluctuations
 in the absence of gravitational fluctuations.
We found that the
statistical net baryon number
 fluctuations are large at the QCD phase transition at $\mu_B \sim 0$
 and at T $\sim$ 10 MeV and less and at high $\mu_B$. 
These are regions where such  stars could be formed.
\noindent
The mass dominating part of the hadron shell is very similar to a neutron star.
This can be seen in our estimate of the mass : $\sim \ 1.34$ solar masses,
slightly below the Chandrasekhar bound. The distinctive 
feature to a neutron star
is that the hadron shell does not extend to the center of the 'star'.
\noindent
The quark core taken for itself and according to Tolmans universal solution,
with $\rho_e \ \propto \ r^{-2}$ towards the center is {\em not} stable.
This means that it contains less nucleons {\em counted by number} than
{\em counted by mass}. Yet this does not imply any instability of the 'star'
as a whole, as it would be the case for a one-phase neutron star.
\noindent
The phase boundary between quark core and hadron shell with its surface pressure
stabilizes the 'star'. 
\noindent
Despite genuine baryonic origin, besides the (anti)leptonic parts,
such 'stars' are indeed candidates for {\em any kind} of dark matter.
If 1.8 solar masses indeed characterize the mass of these 'stars' then
they could be observed through gravitational microlensing
of the galactic halo towards the Magellanic clouds  \cite{jetzer}.
However masses much larger than few solar masses are hardly detectable by
lensing.
We also note, that the composition of the 'star' considered here lacks heavy 
elements
and hence the metallicity nearby is unusually low compared to
burning stars of similar mass.
\noindent
They could possibly emit and absorb
H lines, e.g. $\mbox{Lyman}-\alpha$ lines taking into account
gravitational redshift (which is z=0.56 for a mass of 1.8 solar masses)
 and high pressure compared to normal conditions.
\noindent
Another possibility is that they are part
of a gravitating system including one or more shining stars.

\noindent
{\bf Acknowledgments}

\noindent
We thank Dr. J. Schaffner-Bielich, Dr. R. von Steiger
and Prof. K. Pretzl for discussions.


\vspace*{0.1cm}

\begin{thebibliography}{99}
%\vspace*{-0.2cm}

%\cite{astrostrangelets1,astrostars1,astrostars2,astrostars3}.

\bibitem{astrostrangelets1} F. Weber, 
%Strangeness in neutron stars,
		 J. Phys. G27 (2001) 465, astro-ph/0008376,
%\bibitem{astrostars1} 
H. Huber et al.,
%, F. Weber and M.K. Weigel, 
%Symmetric and asymmetric nuclear matter in the relativsitic approach at finite temperatures,
	  Phys. Rev. C57 (1998) 3484, nucl-th/9803026,
%
%\bibitem{astrostars2} 
J. Madsen, 
% Probing strange stars and color superconductivity by 
% R mode instabilities in millisecond pulsars,
     Phys. Rev. Lett. 85 (2000) 10, astro-ph/9912418,
% J. Madsen, 
% Physics and astrophysics of strange quark matter, 
astro-ph/9809032,
%in 'Hadrons in dense matter and hadrosynthesis', Lecture Notes in Physics, Springer Verlag (ed. J.  Cleymans), p. 162,  Cape Town 1998. 
%
%
%\bibitem{astrostars3}
 J. A. Pons et  al., 
% F. M. Walter, J. M. Lattimer, M. Prakash, R.  Neuhaeuser and P. An,
% Towards a mass and radius determination of the nearby isolated neutron star RX J185635-3754,
astro-ph/0107404,
%
%
M. Prakash et al., 
% J. M. Lattimer, J. A. Pons, A. W. Steiner and S. Reddy, 
astro-ph/0012136, 
% to appear in the proceedings of ECT* International Workshop on Physics of Neutron Star Interiors
% (NSI00), Trento, Italy, 19 Jun - 7 Jul 2000. 
%
%
%
J.M. Lattimer and M. Prakash,
% Neutron star structure and the equation of state,
Astrophys. J. 550 (2001) 426, astro-ph/0002232. 

\bibitem{tolman} R.C. Tolman, Phys. Rev. 55 (1939) 364.

\bibitem{issi} P. Minkowski, S. Kabana,
proceedings of the workshop on 'Matter in
the universe', ISSI, Bern, March 2001, to
appear in Space Science Reviews, Kluwer Acedemic Publishers,
the Netherlands, at press.


% \bibitem{straumann} 
% N. Straumann, Allgeneine Relativitaetstheorie  und relativistische
% Astrophysik.
% page 349 

%bibitem{tov} N. Straumann, Allgeneine Relativitaetstheorie  und relativistische
%Astrophysik, page 334, equation (14).


\bibitem{mapping} S. Kabana and P. Minkowski, New. J. Phys. 3 (2001) 4.

\bibitem{bogolubov} N. N. Bogolubov, Journal of Physics USSR, 
XI (1947) 23.

\bibitem{jetzer} Ph. Jetzer, astro-ph/9901058, proceedings of EC-TMR Euroconference on 3K Cosmology,
Rome, Italy, 5-10 Oct 1998. 

\bibitem{na52}
% badhonnef
%G. Ambrosini et al., (NA52 coll.): nucl-ex/0011016,
K. Pretzl et al. (NA52 Collaboration), Proc. of the 241. 
WE-Heraeus Seminar : Symposium on Fundamental Issues in
Elementary Matter, Bad Honnef, Germany 25.-29. September 2000,
published in Acta Physica Hungarica 2001, pp. 277-287, nucl-ex/0011016, \\
% baryon paper
G. Ambrosini et al. (NA52 Collaboration),
Phys. Lett. B 417 (1998) 202, \\
% centr
G. Ambrosini et al. (NA52 Collaboration),
New J. of Phys. 1 (1999) 22, \\
% na35
T. Alber et al. Phys. Lett. B 366 (1996) 56, \\
% star
C. Adler et al. (STAR Collaboration), 
Phys. Rev. Lett. 87 (2001) 262301 and 279902,
nucl-ex/0108022.

\end{thebibliography}
\end{document}